\documentclass[pra,superscriptaddress, amsmath,amssymb,aps,floatfix,]{revtex4-1}

\usepackage{graphicx}
\usepackage{dcolumn}
\usepackage{bm}
\usepackage{hyperref}
\usepackage{gensymb}
\usepackage{siunitx}

\begin{document}

\title{Microcavity Polaritons for Quantum simulation}

\author{Thomas Boulier}
\affiliation{Laboratoire Kastler Brossel, Sorbonne Universit\'{e}, CNRS, ENS-Universit\'{e} PSL, Coll\`{e}ge de France, Paris 75005, France}
\author{Maxime J. Jacquet}
\affiliation{Laboratoire Kastler Brossel, Sorbonne Universit\'{e}, CNRS, ENS-Universit\'{e} PSL, Coll\`{e}ge de France, Paris 75005, France}
\author{Anne Ma\^itre}
\affiliation{Laboratoire Kastler Brossel, Sorbonne Universit\'{e}, CNRS, ENS-Universit\'{e} PSL, Coll\`{e}ge de France, Paris 75005, France}
\author{Giovanni Lerario}
\affiliation{Laboratoire Kastler Brossel, Sorbonne Universit\'{e}, CNRS, ENS-Universit\'{e} PSL, Coll\`{e}ge de France, Paris 75005, France}
\author{Ferdinand Claude}
\affiliation{Laboratoire Kastler Brossel, Sorbonne Universit\'{e}, CNRS, ENS-Universit\'{e} PSL, Coll\`{e}ge de France, Paris 75005, France}
\author{Simon Pigeon}
\affiliation{Laboratoire Kastler Brossel, Sorbonne Universit\'{e}, CNRS, ENS-Universit\'{e} PSL, Coll\`{e}ge de France, Paris 75005, France}
\author{Quentin Glorieux}
\affiliation{Laboratoire Kastler Brossel, Sorbonne Universit\'{e}, CNRS, ENS-Universit\'{e} PSL, Coll\`{e}ge de France, Paris 75005, France}
\author{Alberto Bramati}
\affiliation{Laboratoire Kastler Brossel, Sorbonne Universit\'{e}, CNRS, ENS-Universit\'{e} PSL, Coll\`{e}ge de France, Paris 75005, France}
\author{Elisabeth Giacobino}
\email{elisabeth.giacobino@lkb.upmc.fr}
\affiliation{Laboratoire Kastler Brossel, Sorbonne Universit\'{e}, CNRS, ENS-Universit\'{e} PSL, Coll\`{e}ge de France, Paris 75005, France}
\author{Alberto Amo}
\affiliation{Univ. Lille, CNRS, UMR 8523, Laboratoire de Physique des Lasers Atomes et Mol\'{e}cules (PhLAM), F-59000 Lille, France}
\author{Jacqueline Bloch}
\email{jacqueline.bloch@c2n.upsaclay.fr}
\affiliation{Universit\'{e} Paris-Saclay, CNRS, Centre de Nanosciences et de Nanotechnologies, 91120 Palaiseau, France}

\date{\today}

\begin{abstract}
Quantum simulations are one of the pillars of quantum technologies. 
These simulations provide insight in fields as varied as high energy physics, many-body physics, or cosmology to name only a few. 
Several platforms, ranging from ultracold-atoms to superconducting circuits through trapped ions have been proposed as quantum simulators.
This article reviews recent developments in another well established platform for quantum simulations: polaritons in semiconductor microcavities.
These quasiparticles obey a nonlinear Schr\"odigner equation (NLSE), and their propagation in the medium can be understood in terms of quantum hydrodynamics. As such, they are considered as ``fluids of light''.
The challenge of quantum simulations is the engineering of configurations in which the potential energy and the nonlinear interactions in the NLSE can be controlled.
Here, we revisit some landmark experiments with polaritons in microcavities, discuss how the various properties of these systems may be used in quantum simulations, and highlight the richness of polariton systems to explore non-equilibrium physics.
\end{abstract}

\maketitle

\section{Introduction}

Quantum simulations are a very active line of research within the large family of quantum technologies.
These simulations broadly aim at reproducing and experimentally predicting the behaviour of complex quantum systems that are difficult to model analytically or numerically \cite{Georgescu_quantsim_2014}.
Typically, this can be the study of the interactions of a ensemble of quantum particles such as bosons \cite{Bloch_review_2008} and/or the intrinsically non-equilibrium dynamics of open systems.
That is, systems for which calculating or computing the wave function is complex or computationally heavy.
Among the numerous possible platforms, cold gases of atoms or molecules confined in traps, on atom chips or in optical lattices have historically attracted a lot of interest because of their programmability and their strong long-range interactions.
The latter feature is crucial to model a variety of many-body systems and phenomena, such as quantum phase transitions (\textit{e.g.} superfluid to Mott insulator transition) and spin models (\textit{e.g.} Ising or Heisenberg chains), which play an important role in our fundamental understanding of complex quantum systems.
The programmability of quantum simulators implies a high level of control on the system, which should range from the single particle level to a large number of particles so as to approach the thermodynamic limit.
Photonic simulators are extremely promising in this regard.
These systems rely on the propagation of light in a material with a strong non-linearity to engineer an effective photon-photon interaction in integrated platforms based on nanotechnologies.
Polariton condensates relying on cavity quantum electrodynamics in semiconductor micro- or nano-structures also constitute a compelling simulation platform thanks to its integrability and scalability.
As will be shown in this article, the polariton platform is one of the prominent architectures closest to reaching maturity.

Quantum simulations based on photonic systems have the major advantage of being relatively easy to engineer. This can be achieved in an optical cavity containing a material medium interacting with light. The dynamics of light are then modified by the non-linearity of the material system. In a semiconductor microcavity, photons can populate excited electronic states called excitons. In the strong coupling regime arising between excitons and photons, the eigenstates are polaritons that are composite bosons; half-light, half-matter particles~\cite{weisbuch1992observation, kavokin}. 
The photonic part of these quasiparticle allows their optical creation and detection, while their excitonic part provide non-linear interactions. At the same time, they acquire an effective mass through propagation in the cavity, and the finite cavity finesse implies a finite particle lifetime.
Typically, a resonant laser excitation creates a coherent polariton flow with an adjustable density and velocity. All the flow parameters such as density, phase, velocity and spatial coherence can then be directly measured from the light escaping the cavity.

In polariton ensembles, effective photon-photon interactions stem from the exciton-exciton Coulomb interactions, resulting in a hydrodynamics-like behavior. Moreover, the finite polariton lifetime allows to easily implement situations where non-equilibrium physics becomes important.
This makes the polariton system a unique platform to simulate quantum fluids displaying rich physics~\cite{Carusotto2013}, and opens the way to simulating astrophysical relativistic properties such as the physics of black holes \cite{volovikUniverseHeliumDroplet2003}.
A potential landscape can be designed by lateral patterning of microcavities \cite{deveaud2006} or by using optical means \cite{AmoPigeon2010, Wertz2010}. This allows to simulate condensed matter physics, including strongly correlated and lattice systems.

These features make the polariton platform a unique venue to simulate various physical effects in a semiconductor chip and to evidence properties very difficult to access in other systems.

This article is structured as follows: in section 2, we review the physics of polaritons in semiconductor microcavities. 
We start from light-matter interactions in the strong coupling regime and introduce the essential equation describing the dynamics of polaritons --- the Gross-Pitaevskii equation.
We then explore the bistable behaviour of the system as a function of the pump beam parameters.
In section 3, we discuss the superfluid properties of polaritons in microcavities.
To this end, we study the evolution of small fluid perturbations.
We review the experiments which first demonstrated polaritons superfluidity, and proceed to present one stricking effect resulting from superfluid properties, namely the formation of vortices and vortex lattices.
In the fourth section, we show how to simulate quantum turbulences in a polariton flow.
We review the physical model of turbulence and show how vortices and solitons appear in dissipative flows.
We also discuss a newly established method to engineer sustained polariton flows where novel soliton behaviors emerge.
Furthermore, we highlight how the driven-dissipative nature of polaritons provides opportunities for unveiling hydrodynamics phenomena that are not accessible in equilibrium quantum fluids.
In section 5, we present the physics of bright solitons --- another type of non-linear excitations that can be resonantly driven in a polariton fluid.
Finally, in section 6 we revisit the potential of analogue gravity physics to explore effects of quantum fields on curved spacetimes with microcavity polaritons.

\section{Microcavity polaritons}

A typical polariton platform is made of $In_{x}Ga_{1-x}As/GaAs$ semiconductor quantum wells (with $x$ of the order of $0.05$) embedded between two highly reflecting planar Bragg mirrors, separated by a distance of the order of a few wavelengths. Such microcavities have a high finesse, ranging from 5000 to $10^5$. When photons propagate in an optical cavity, the resonance condition with a cavity mode is given by $p\lambda/2 = n_r\,l\cos \theta $, where $p$ is an integer, $\lambda$ is the mode wavelength in vacuum, $n_r$ the cavity layer index of refraction, $l$ the cavity thickness and  $\theta$ the angle of incidence (inside the cavity) with respect to the normal of the cavity. Introducing the wavevector $\mathbf{k}$ inside the cavity, with $k=2\pi n_r/\lambda$, this implies that the component $k_{z}$ of  $\mathbf{k}$ perpendicular to the cavity plane is fixed to $p\pi/l$. In the case of a small angle of incidence (or a small component $k_{\parallel}$ of the wavevector parallel to the cavity plane)  the dispersion can then be written as:

\begin{equation}
\label{eq:disprelpol}
    \omega(k_{\parallel})=\frac{c}{n_r}\sqrt{k_z^2+k_{\parallel}^2}\approx \frac{c}{n_r} k_z +\frac{\hbar k_{\parallel}^2}{2m},
\end{equation}

where $m\equiv \hbar n_r k_z/c$ is the effective photon mass $m$ for motion in the cavity plane. The dispersion is similar to the parabolic dispersion relation of a massive particle in 2D. The mass is very small, on the order of $\sim10^{-5}$ the mass of the electron.

\subsection{Strong coupling}

The cavity photons interact with the active medium, here InGaAs quantum wells with a thickness of about 10 nm and placed at the anti-nodes of the GaAs microcavity electromagnetic field. Photons can excite an electron from the filled valence band to the conduction band, thereby creating a hole in the valence band. The electron-hole system possesses bound states called excitonic states, that couple efficiently to photons with an energy slightly below the bandgap. In a narrow quantum well, excitons states are quantized along the $z$ direction and are free to propagate in the 2D cavity plane. The dispersion curve for the excitons is then parabolic as a function of $k_{\parallel}$, but since the mass of the excitons is much higher than the effective mass of a photon their dispersion curve is flat at small $k_{\parallel}$.

\begin{figure}[ht]
\begin{center}
  \includegraphics[width=0.7\linewidth]{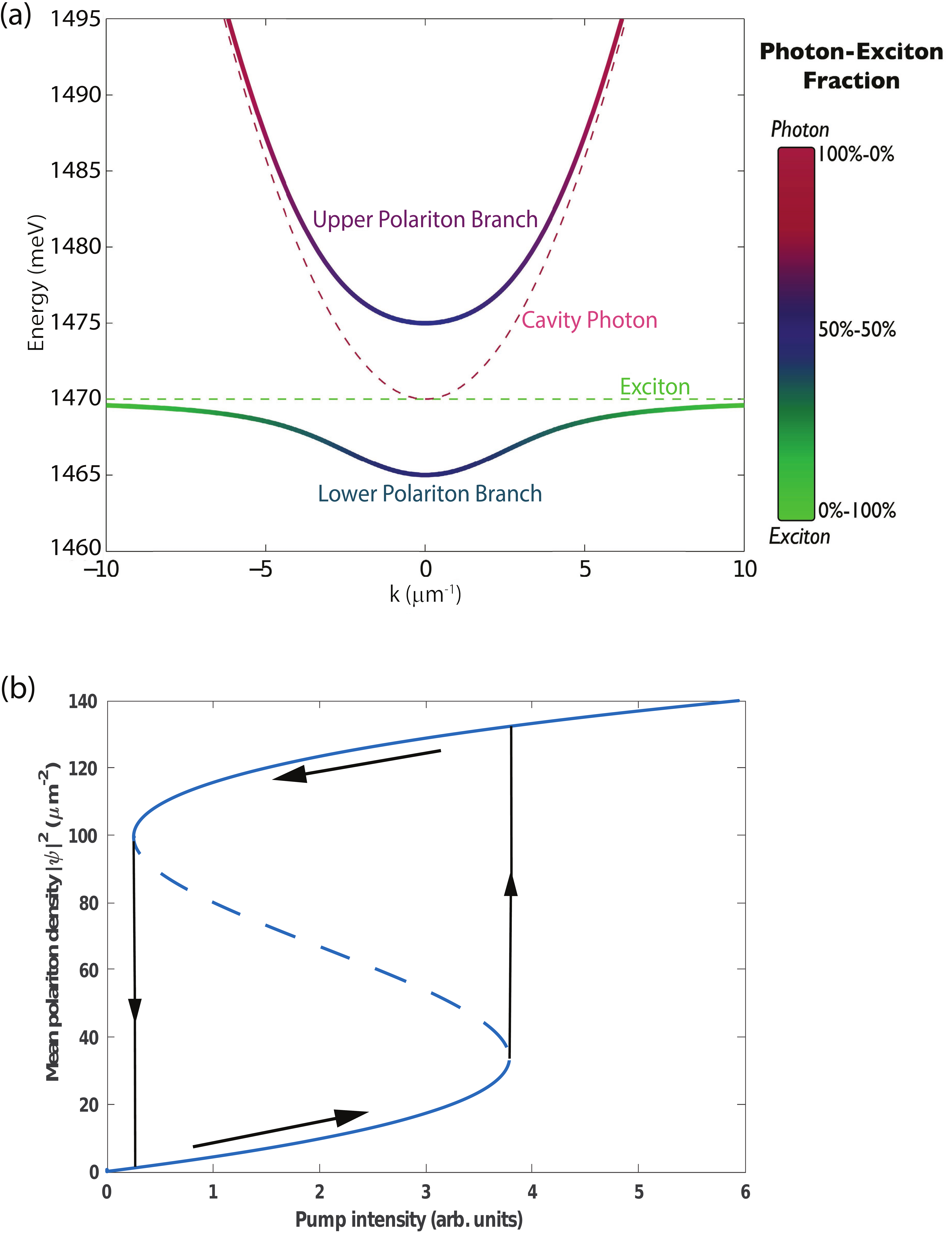}
  \caption{(a) Dotted lines: dispersion curves of the exciton and microcavity photons in the absence of coupling; full lines dispersion curves of the lower and upper polaritons, in the presence of strong coupling. (b) Bistability curve giving the mean number of polaritons versus the input laser intensity. In the bistable region, the dashed line is the unstable branch. The arrows indicate the hysteresis cycle obtained by scanning the input power in the two directions.}
  \label{fig1}
\end{center}
\end{figure}

The light-matter coupling can be represented by the Hamiltonien:

\begin{equation}
\label{eq:hamil}
   \mathbf{H_{k_{\parallel}}}=\hbar \omega(k_{\parallel}) \mathbf{a^{\dag}} \mathbf{a} + E_{exc}\mathbf{b^{\dag}}\mathbf{b} + \frac{\hbar \Omega_{R}}{2}(\mathbf{a^{\dag}}\mathbf{b} + \mathbf{b^{\dag}}\mathbf{a})
\end{equation}

where $\textbf{a}$ (resp. $\textbf{b}$) denotes the anihilation operator of a photon (resp. an exciton). In this kind of microcavity, the light-matter interaction represented by the Rabi frequency $\Omega_{R}$ is much larger than the photons and excitons decay rates, and the strong coupling regime takes place \cite{weisbuch1992observation}. The eigenstates are linear superpositions of photon and exciton states, known as cavity polaritons. Their decay rate is the weighted average of the cavity  and exciton decay rates. The two eigenstates, called the upper and the lower polariton branches, are separated by an energy of the order of the Rabi splitting as shown in Figure~\ref{fig1}(a). Their dispersion curves depend on the photon and the exciton dispersion and in particular, a curvature appears on the lower polariton branch with an effective mass comparable to the photon one. These polaritons are quite interesting objects: they are a superposition of essentially motionless matter particles (the excitons) and of photons that propagate in the cavity, "jumping" from an exciton to the next. The state of the system can be accessed experimentally thanks to the continuous escape of photons out of the cavity.

\subsection{Nonlinear properties}

Coulomb interaction between the carriers gives rise to an effective exciton-exciton interaction. Since this interaction is small compared to the Rabi splitting, one can neglect the nonlinear interaction between the upper and lower polariton branches, and it can be projected on the polariton basis, yielding an effective polariton-polariton interaction represented by a Kerr-like nonlinear term. Consequently, the evolution of the lower polariton wavefunction can be described by the following nonlinear Schr\"odinger equation

\begin{equation}
    \label{eq:GPE}
    i\hbar\frac{\partial\psi_{LP}}{\partial t}=(\hbar\omega_{LP}  -\frac{\hbar^2}{2m}\nabla^2+ \hbar g|\psi_{LP}|^2 -i\hbar\gamma)\psi_{LP} + i\hbar\eta F_0.
\end{equation}

The first term on the right-hand side represents the ground energy of the lower polariton state $\hbar\omega_{LP}$; the second term is the kinetic energy in the cavity plane, with $m$ the effective mass for the lower polariton branch; the third term is the nonlinear interaction $\hbar g$, which is positive, meaning that the interaction is repulsive; the fourth and fifth terms are respectively the losses due to the polariton decay rate $\gamma$ and the polariton source from the pump laser field of amplitude $\hbar F_0$ which generates polaritons with efficiency $\eta$.

These composite bosons (spin $\pm 1$) possess interesting properties due to their peculiar dispersion relation and their efficient mutual interaction. When operated in liquid helium at 4K, and assuming thermal equilibrium, their small effective mass implies a large coherence length

\begin{equation}\label{coherencelength}
   \lambda_{T} = \sqrt{\frac{2\pi\hbar^{2}}{mk_{B}T}}
\end{equation}

of the order of 1 to 2 $\mu m$, while the mean distance between polaritons is typically of the order of 0.1$\mu m$ in standard experimental conditions. This is a key ingredient for Bose-Einstein condensation~\cite{kasprzak2006bose} and it enables the building of various many-body coherent quantum effects, which open promising avenues for quantum simulation.

\subsection{Bistability}
\label{sec:bistability}

We now look at the system behaviour as a function of the pump laser parameters. When pumped by a quasi-resonant laser in a nearly plane wave geometry, the polariton wavefunction has the same frequency as the laser and one can look for solutions of the form

\begin{equation}\label{wavefunction}
    \psi_{LP} = \psi_{0} e^{ik_{0}r} e^{-i\omega_{0}t},
\end{equation}

where $\omega_{0}$ and $k_{0}$ are the frequency of the laser and its wave number in the cavity plane.
Using Equation (\ref{eq:GPE}), the polariton wave function is given by:

\begin{equation}
    \label{eq:GPEsol}
    (\omega_{0} - \omega_{LP}  -\frac{\hbar}{2m}k_{0}^2 - g|\psi_{0}|^2 +i\gamma)\psi_{0} = i\eta F_0.
\end{equation}

Setting $ \omega_{LP}  +\frac{\hbar}{2m}k_{0}^2 -\omega_{0} = \delta $, $|\psi_{0}|^2 =n $ and $|F_0|^{2} = I_0 $ , we multiply the previous equation by its complex conjugate to obtain the mean number of polaritons

\begin{equation}
    \label{eq:intensity}
    n [(\delta + gn)^{2} +\gamma^{2}] = \eta^{2} I_0.
\end{equation}

The polariton density $n$ shows a bistable behavior as a function of the laser excitation power \cite{baas2004optical} as shown in Figure~\ref{fig1}(b). The bistable behavior is obtained for $\delta<0$ and $\delta^{2}>3\gamma^{2}$ . The bistability turning points can be shown to be

\begin{equation}
    \label{eq:turning points}
    g n = - \frac{2}{3}\delta \pm\frac{1}{3}\sqrt{\delta^{2}-3\gamma^{2}}
\end{equation}

We see that the upper bistability turning point (on the left hand side) is close to $gn=-\delta$ (if $\delta\gg\gamma$), that is at the point where the nonlinear interaction compensates the polariton-laser detuning. These bistable properties will be shown to bring interesting properties to the system.

\section{Superfluid properties of polaritons in a microcavity}

The evolution of the lower polariton wavefunction is described by the non-linear Schrödinger equation (\ref{eq:GPE}) similar to the Gross-Pitaevskii equation describing Bose-Einstein condensates or quantum fluids at zero temperature, which makes polaritons very promising for simulating quantum fluids.
One of the main differences with respect to conservative systems comes from the last two terms representing the losses due to the decay rate of the polaritons and the polariton source from the pump laser.
They are not present in the usual Gross-Pitaevskii equation and here they account for the driven-dissipative nature of the cavity polariton system.

Similar to their conservative counterpart, polariton fluids were predicted to show collective phenomena such as Bose Einstein condensation or superfluidity~\cite{chiao1999bogoliubov,carusotto2004probing,ciuti2005quantum} and these properties have been observed in several groups in the past years~\cite{kasprzak2006bose, amo2009superfluidity,lagoudakis2008quantized,amo2009collective,utsunomiya2008observation}.

\subsection{Bogoliubov dispersion}

In order to investigate the superfluid properties of polaritons, we will consider the Landau criterion usually applied to superfluids in conservative systems. This criterion establishes the maximal speed of the quantum fluid, up to which no density excitation can be created in the fluid when encountering an obstacle. Using Equation (\ref{eq:GPE}) we study the onset of the collective mechanisms in polariton fluids, which appear when the nonlinear interactions become significant. These properties can be studied by investigating the evolution of small perturbations. This is done using the Bogoliubov method, i.e. linearizing the Gross-Pitaevskii equation in the vicinity of a working point:

\begin{equation}\label{eq:perturbedwf}
    \psi_{LP}(r,t) = \psi_{LP}^0(r,t) + \delta\psi(r,t)
\end{equation}

We first assume a coherent pump with a frequency $\omega_0$ and a wave number in the cavity plane $k_0=0$.  The steady state is then $\psi_{LP}^0= \psi_{0}e^{i\omega_0 t}$ and $\delta\psi(r,t)$ is given by the equation   

\begin{equation}\label{deltapsi}
    i\hbar\frac{\partial}{\partial t}\left(
       \begin{array}{c}
          \delta\psi(r,t)\\
          \delta\psi^{*}(r,t)\\
       \end{array}
     \right) = \hbar \textsl{L}_{Bog} \left(
       \begin{array}{c}
          \delta\psi(r,t)\\
          \delta\psi^{*}(r,t)\\
       \end{array}
       \right)
\end{equation}

The Bogoliubov operator $\textsl{L}_{Bog}$ can be written as

\begin{equation}\label{bogoliubov}
   \textsl{L}_{Bog}=\left(
                      \begin{array}{cc}
                        \frac{\hbar k^2}{2m}+ \Delta + 2gn -i\gamma & gn \\
                        -gn & -\frac{\hbar k^2}{2m}- \Delta - 2gn -i\gamma \\
                      \end{array}
                    \right)
\end{equation}

where $k$ is the wave number of the perturbation, $\Delta=\omega_{LP}-\omega_0$, and  $n=|\psi_0|^2$ is the polariton density.
A particularly interesting regime appears in the bistable case when $\Delta=-gn$, which corresponds to the turning point of the bistability upper branch (Figure~\ref{fig1}(b). In that case, the nonlinear shift $gn$ compensates for the laser detuning $\Delta$ and the resulting Bogoliubov dispersion is

\begin{equation}\label{Bog_disp}
    \omega_B(k) = \pm\sqrt{\frac{\hbar k^2}{2m}(\frac{\hbar k^2}{2m}+2gn)} - i\gamma.
\end{equation}

Very different behaviors can occur depending on how the perturbation wavelength compares with the healing length $\xi=\sqrt{\hbar/mgn}$, which is the minimum distance for changes in the polariton density due to particle interactions. Concentrating on the real part, when $k$ is small compared to the inverse healing length, a sound-like dispersion occurs:

\begin{equation}\label{sonic}
    \omega_B(k) = \pm\ c_s k
\end{equation}

where the speed of sound is $c_s=\sqrt{\hbar gn/m}$. A parabolic dispersion is recovered for $k \gg 1/\xi$,

\begin{equation}\label{parabolic}
    \omega_B(k) = \pm(\frac{\hbar k^2}{2m}+ gn).
\end{equation}

If the perturbation is occurring in a moving polariton flow ($k_0\neq 0$) with velocity $v_0$, the excitation spectrum written in the laboratory reference frame is

\begin{equation}\label{Bog_disp_2}
    \omega_B(\mathbf{k}) = \mathbf{(k-k_0)}.\mathbf{v_0} \pm\sqrt{\frac{\hbar(\mathbf{k-k_0})^2}{2m}(\frac{\hbar(\mathbf{k-k_0})^2}{2m} +2gn)}.
\end{equation}

In this case the condition established above for the nonlinear shift to compensate the polariton-laser detuning, i.e. for operating near the bistability upper turning point, is $ \Delta = \omega_{LP}  +\frac{\hbar}{2m}k_{0}^2 -\omega_{0} = - gn $

\begin{figure}[ht]
\begin{center}
  \includegraphics[width=0.6\linewidth]{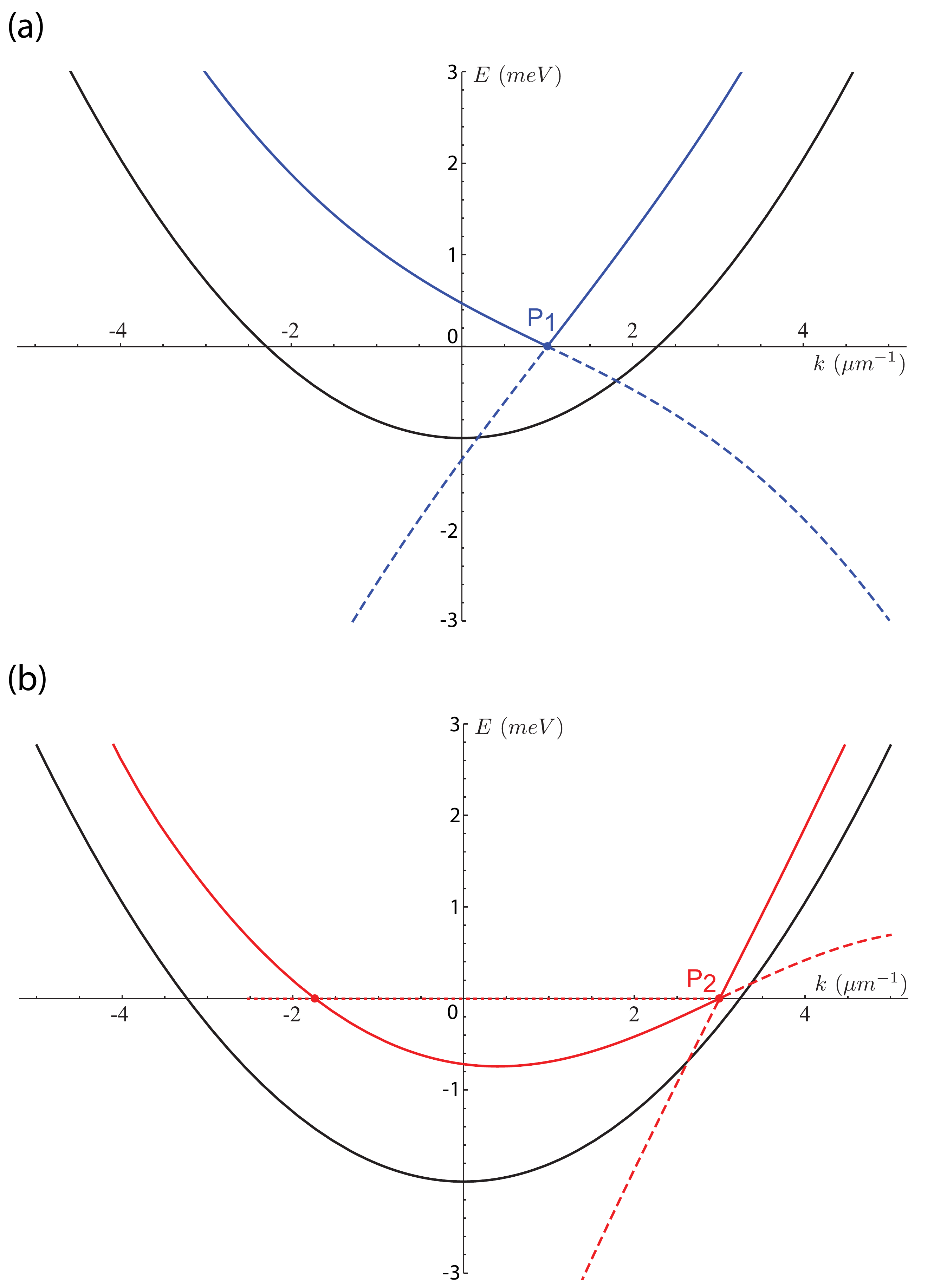}
  \caption{Dispersion curves for small perturbation in the polariton fluid (Bogoliubov modes). The dots marked $P_1$ and $P_2$ indicate the position of the (quasi) resonant pumping. In both instances (a-b), the black curve represents the unmodified dispersion relation obtained at small polariton density (negligible interactions), while the colored curves represents the modified dispersion in presence of interactions. (a) low group velocity of the fluid (subsonic), showing a linear sonic dispersion of the perturbation. The injected polaritons cannot scatter due to the absence of any other state at the pump energy, while in (b), supersonic velocity, the dispersion is also linear around the pump energy, but there are available states for scattering.}
  \label{fig2}
\end{center}
\end{figure}

Figure~\ref{fig2}(a) shows the change occurring in the dispersion when one goes from a linear case (negligible nonlinear interactions) to the case given above, where the fluid velocity $v_0$ is smaller than the sound velocity $c_s$ (subsonic case).
 
 \begin{equation}\label{sonic_2}
    \omega_B(\mathbf{k}) = \mathbf{(k-k_0)}.\mathbf{v_0} \pm c_s|\mathbf{k-k_0}| 
\end{equation}

The point $k = k_0$ corresponds to the minimum of the dispersion curve as can be seen in Figure~\ref{fig2}(a), as the slopes on each side have opposite signs. This gives rise to superfluidity since elastic scattering (for example against an obstacle) is not possible in this case, thus fulfilling the Landau criterion.

When the fluid is supersonic ($v_0 > cs$), as shown in Figure~\ref{fig2}(b), the dispersion curve still has a linear dispersion in the vicinity of the perturbation, but the slopes on each side of $k_0$ have the same sign. The dispersion curve minimum is therefore lower, and scattering on a defect is allowed.

\subsection{Probing the Landau criterion}
 
\begin{figure}[ht]
\begin{center}
  \includegraphics[width=0.75\linewidth]{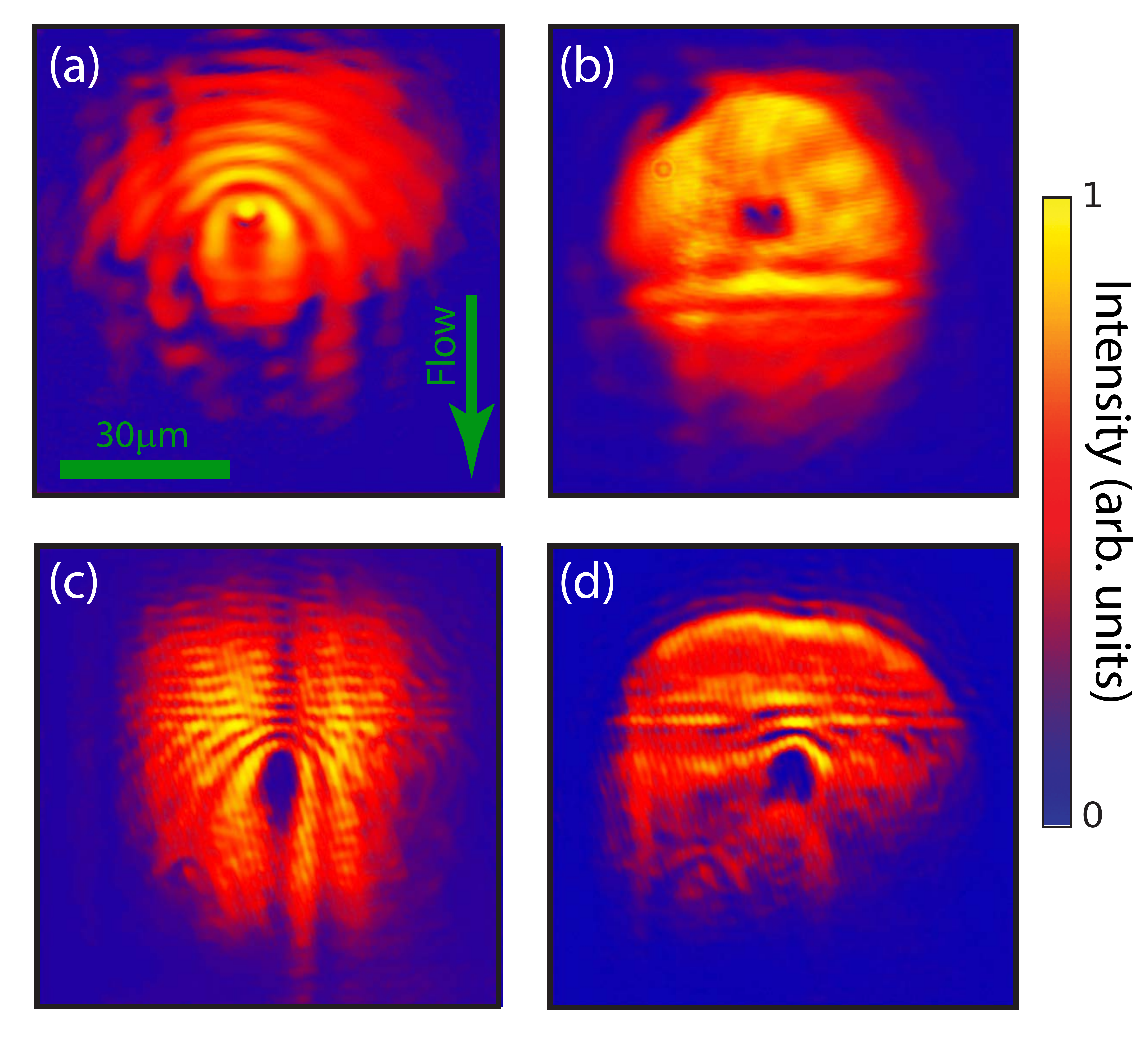}
  \caption{Experimental images of the intensity transmitted through the microcavity. (a) and (c): low density case, where interference fringes can be seen between the incoming and scattered flow. (b): in the high density, subsonic case there is no scattering, demonstrating superfluidity. (d): high density, supersonic case where scattering leads to Cerenkov linear wavefronts. Reproduced with permission, from~\cite{amo2009superfluidity}. Copyright 2009, Nature Publishing Group.}
  \label{fig3}
\end{center}
\end{figure}

These superfluid properties were studied experimentally~\cite{amo2009superfluidity} with a semiconductor microcavity as described above. A laser with a frequency close to the cavity resonance was focused on the cavity with a variable incidence angle. The excitation spot was placed on a defect present in the sample. The in-plane fluid velocity was changed varying the pump laser angle of incidence on the sample. First experiments were performed at a small pump angle, $2.6\degree$, corresponding to $k_0 = 0.34 \mu m^{-1}$ and to a fluid group velocity $v_0=(\hbar k_0)/m=0.64 \mu m/ps$ (Figure~\ref{fig3}), upper panel). At low excitation power, the fluid density is low and the interactions are negligible: polaritons are elastically scattered by the defect, generating cylindrical waves propagating away. This gives parabolic interference fringes with the incident wave, as can be seen in Figure~\ref{fig3}(a). When the laser intensity is increased up to the value where $gn\sim \Delta$, polariton-polariton interactions increase and cause a dramatic change in the polariton dispersion in the vicinity of the defect. In Figure~\ref{fig3}(b) the sound velocity reaches $0.8\mu m/ps > v_0$ due to the large interactions, for which there is no more possibility for elastic scattering. The interference fringes disappear, demonstrating the emergence of a friction-less flow characteristic of the superfluid regime.

Next the incidence angle was increased to $4\degree$, corresponding to a wave number of $0.52 \mu m^{-1}$ and a fluid velocity $v_0 = 0.98 \mu m/ps > c_s$ (Figure \ref{fig3}, lower panel). At low power, a parabolic-shaped modulation is observed as in the case above. At higher density, polariton waves are still scattered by the defect as predicted in the supersonic case. Interestingly, the shape of the interference fringes is modified, displaying the straight lines characteristic Cerenkov wavefronts. From the angle $\varphi$ between the linear fringes one can measure the sound velocity $\sin\varphi/2=c_s/v_0$, which confirms the value mentioned above, $c_s = 0.8\mu m/ps$.

These experiments show that excitations are energetically forbidden in a polariton condensate flowing at subsonic speeds: the gas will remain superfluid with an unperturbed flow around obstacles. On the other hand, at supersonic speeds the presence of the obstacle induces dissipation (drag) in the form of density modulations which radiate away (downstream) from the potential barrier in the form of linear shock waves. Thanks to the remarkable similarity between polariton fluids of light and material fluids, superfluidity, the extraordinary property whereby interacting Bose-Einstein condensates flow without friction, can be achieved with polaritons. This is evidenced by the absence of excitation when polariton fluids encounter a static obstacle, only occurring when flowing at speeds below the critical velocity.

As in matter superfluids, when the obstacle is much smaller than the healing length of the fluid (the minimum length-scale for density variations in the condensate due to repulsive particle interactions), the critical flow speed for the onset of excitations is given by the Landau criterion and corresponds to the local speed of sound. When the size of the obstacle is larger a richer variety of excitations can be produced, as presented in the next section. The Landau criterion originally developed for conservative systems provides a powerful understanding of the back-scattering suppression shown in Figure~\ref{fig3}. The driven-dissipative nature of the polariton fluid results in additional effects with no analogue in conservative systems, such as intricate parametric effects resulting in the emergence of a Goldstone mode~\cite{Wouters2007b} or in negative drag forces~\cite{VanRegemortel2014,Carusotto2013}.
	
\subsection{Formation of lattices of vortices}

An important property of quantum fluids is their irrotational character. This was demonstrated in atomic Bose-Einstein condensates by injecting angular momentum through stirring of the fluid with a moving obstacle (in this case, a laser)~\cite{madison2000vortex,inouye2001observation}. In early polariton experiments, the irrotational character of the polariton superfluid was demonstrated by the spontaneous formation of vortices in polariton condensation experiments~\cite{lagoudakis2008quantized,lagoudakis2011probing}, and via phase imprinting of vortices~\cite{sanvitto2010persistent,krizhanovskii2010effect} up to orbital angular momentum $L~=~2$.  

More recently, higher orbital angular momenta have been injected in a polariton ensemble with the use of four laser beams arranged in a square and injecting polaritons propagating towards the square center~\cite{boulier2016injection}. By tilting the in-plane pumping direction of the laser beams, the four convergent polariton flows can be made to propagate with a small angle relative to the direction to the center, therefore bringing a controlled angular momentum into the polariton fluid. Here the pumps are tilted by an angle  $\phi=21\degree$, which gives an orbital angular momentum $L = 4$.

\begin{figure}[ht]
\begin{center}
  \includegraphics[width=0.9\linewidth]{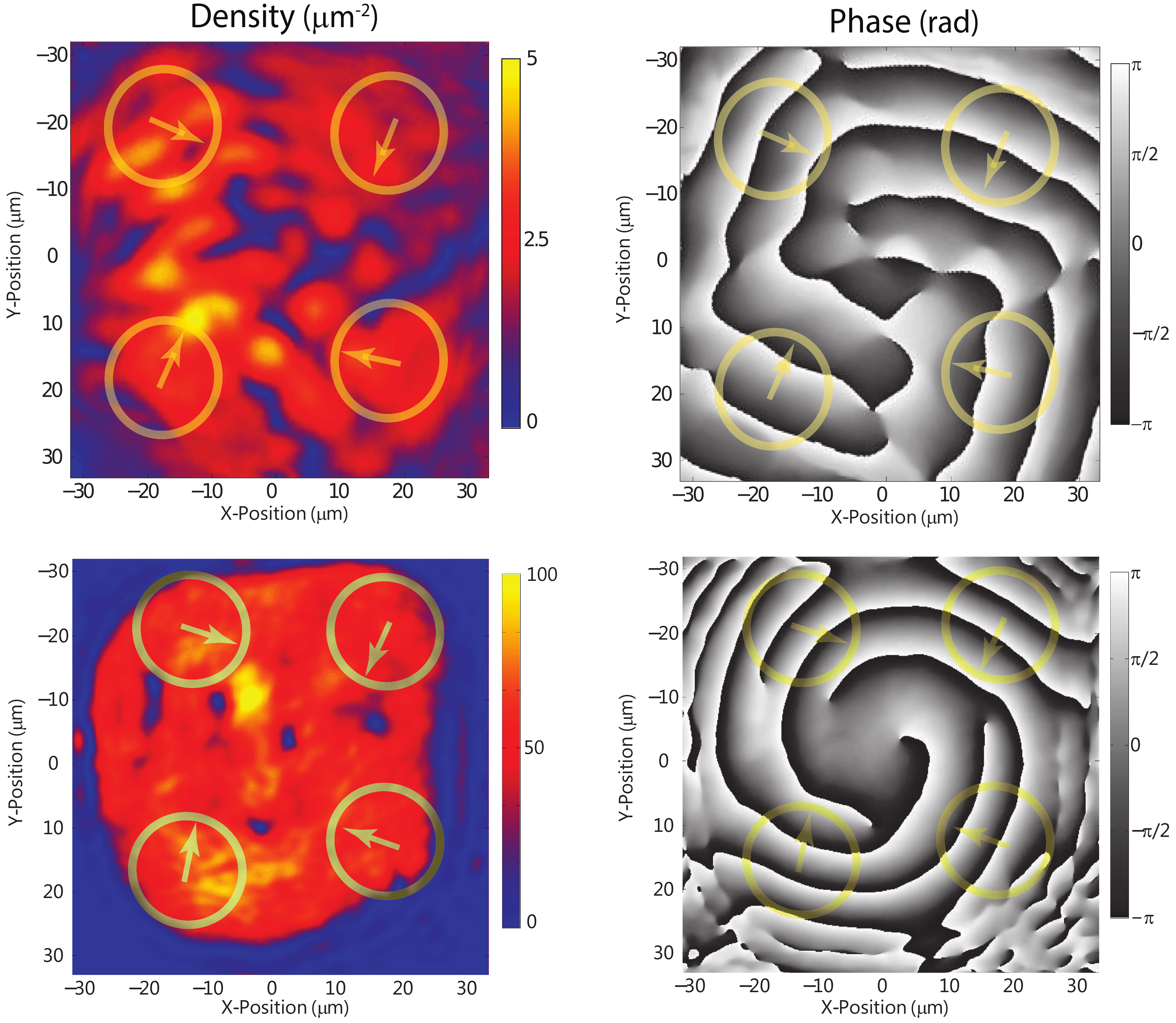}
  \caption{Images of the polariton fluid excited by 4 laser beams (positions and directions indicated by yellow circles and arrows) with angles $\theta =3.5\degree$ and $\phi =21\degree$, yielding an orbital angular momentum of $ L = 4$.  Density (left) and phase (right) maps for low density (upper row) and high density (lower row). In the low density regime an interference pattern is visible with many phase singularities of both signs. In the high density (superfluid) regime, four vortices of the same sign are visible as density holes (left) and as phase singularities (right) near the system center. Reproduced with permission, from ~\cite{boulier2016injection}. Copyright 2016, American Physical Society. }
  \label{fig4}
\end{center}
\end{figure}

At low densities, as shown in Figure \ref{fig4}, a square interference pattern is visible and phase singularities of both signs are visible, with an unequal number of singularities of opposite signs. The difference between the number of vortices (V) and antivortices (AV) is equal to the integer part of the angular momentum $L$ expected from the injected angular momentum. Increasing the polariton density to the superfluid regime, the interference pattern disappears~\cite{boulier2018coherent} and all V-AV pairs annihilate~\cite{cancellieri2014merging}. The conservation of angular momentum results in the presence of elementary vortices of the same sign remaining in the superfluid. Their size is of the order of the healing length (about $1\mu m$) that can be unambiguously defined. Up to five vortices were observed without any antivortex.
This method shows that it should be possible to imprint large value of orbital angular momentum, and to observe many vortices. This opens the way to the study of vortex-vortex interactions and spontaneous vortex lattice formation.

\section{Simulating quantum turbulence in a polariton flow}

The previous experiments show the potential of polariton superfluids for investigation of quantum hydrodynamics. The velocity and the density of the quantum fluid can be tuned at will with the laser excitation, and the polariton density and phase can be measured in real time probing the light emitted by the cavity.

\subsection{Model for turbulence}
 
When the size of the potential barrier is larger than the healing length, the break-up of superfluidity takes place at velocities much lower than the speed of sound far from the barrier. With a large obstacle, different regimes have been observed ranging from superfluidity to the turbulent emission of trains of vortices and the formation of pairs of oblique dark solitons. This is due to the gradient of flow speed occurring around the obstacle, which can be locally higher than the critical speed. The drag force created by the obstacle results in the emission of vortices, at background speeds below the Landau critical velocity for phonon emission, as first predicted in ~\cite{frisch1992transition} for matter superfluids.
 
Vortices are the basic elements of turbulence that is observed in usual fluids, as well as in superfluids. Actually, similar equations can model classical as well as quantum fluids. Using the Madelung transformation, with $\psi = \sqrt{\rho} e^{i\phi}$ ($\phi$ being the phase of the field), one can transform the Gross-Pitaevskii equation into fluid-like equations, i.e. the hydrodynamic continuity and Euler equations
 
 \begin{align}
\label{eq:hydroGPE}
  \frac{\partial \rho}{\partial t} + \mathbf{\nabla}.(\rho \mathbf{v})  = 0, \\
  \frac{\partial \phi}{\partial t} + \frac{m}{2\hbar} v^{2} + g\rho + \frac{\hbar}{2m\rho^{1/2}} \mathbf{\nabla^{2}} (\rho^{1/2}) = 0,
 \end{align}

where the fluid velocity is given by $v=\frac{\hbar}{m}\mathbf{\nabla} \phi$.

Phase singularities play a similar role to vortices in ordinary inviscid fluids. However, while classical fluids described by Euler equations have a continuous vorticity, vortices in quantum fluids are quantized. In the presence of an obstacle larger than the healing length, these equations allow to predict the behaviour of the fluid. For a perfect incompressible flow around a cylindrical obstacle, the local fluid velocity $v$ increases and can be shown to reach twice the initial velocity $v_0$. This induces a local density change that can be calculated for distances not too close to the obstacle using the quantum fluid equations given above in the case of stationary solution. One obtains:

\begin{equation}
   g \rho = g \rho_{0} + \frac{m}{2\hbar}(v_{0}^{2}-v^{2})
\label{velocity}
\end{equation}

This implies that $ c_{s}^{2} - c_{s loc}^{2} =  \frac{1}{2}(v^{2}-v_{0}^{2})$, where $c_{s loc}$ is the local speed of sound. Using the previous equation and taking the local velocity $v$ near the obstacle as equal to twice the initial velocity $v_{0}$~\cite{frisch1992transition}, one can show that $v$ becomes larger than the local speed of sound  $v > c_{s loc}$  when $v_{0}^{2} > \frac{2}{11} c_{s}^{2}$, approximately when $v_{0} > 0.4 c_{s}$. At that point the stationary solution is not valid anymore and local instabilities may appear. This model thus predicts that when the flow speed far upstream of the obstacle is above the revised critical velocity, i.e. $v_{0} > 0.4 c_s$,  vortices and antivortices are released in the wake of the obstacle~\cite{madison2000vortex,josserand1999vortex,winiecki2000vortex,kamchatnov2008stabilization}.

Relying on the similarity of the Gross-Pitaevski equations for matter condensates and polaritons, similar phenomena have been predicted for microcavity polaritons~\cite{pigeon2011hydrodynamic}. Depending on the defect size, the polariton ensemble can either flow unperturbed around the defect, or experience nucleation of vortices and/or solitons at the surface of the defect.
In order not to lock the phase of the flow past the defect, the pump laser must be centered upstream from the defect in such a way that the defect is not illuminated by the laser. The polaritons can then propagate freely outside the pump on a distance which is limited by their lifetime, of the order of $40\mu m$. Along this distance, the fluid phase is free and turbulent polariton flows were demonstrated in the presence of a large obstacle in a microcavity~\cite{amo2011polariton, nardin2011hydrodynamic,Grosso2011, Grosso2012}.

\subsection{Vortices and solitons in dissipative polariton flows}

\begin{figure}
\begin{center}
  \includegraphics[width=0.75\linewidth]{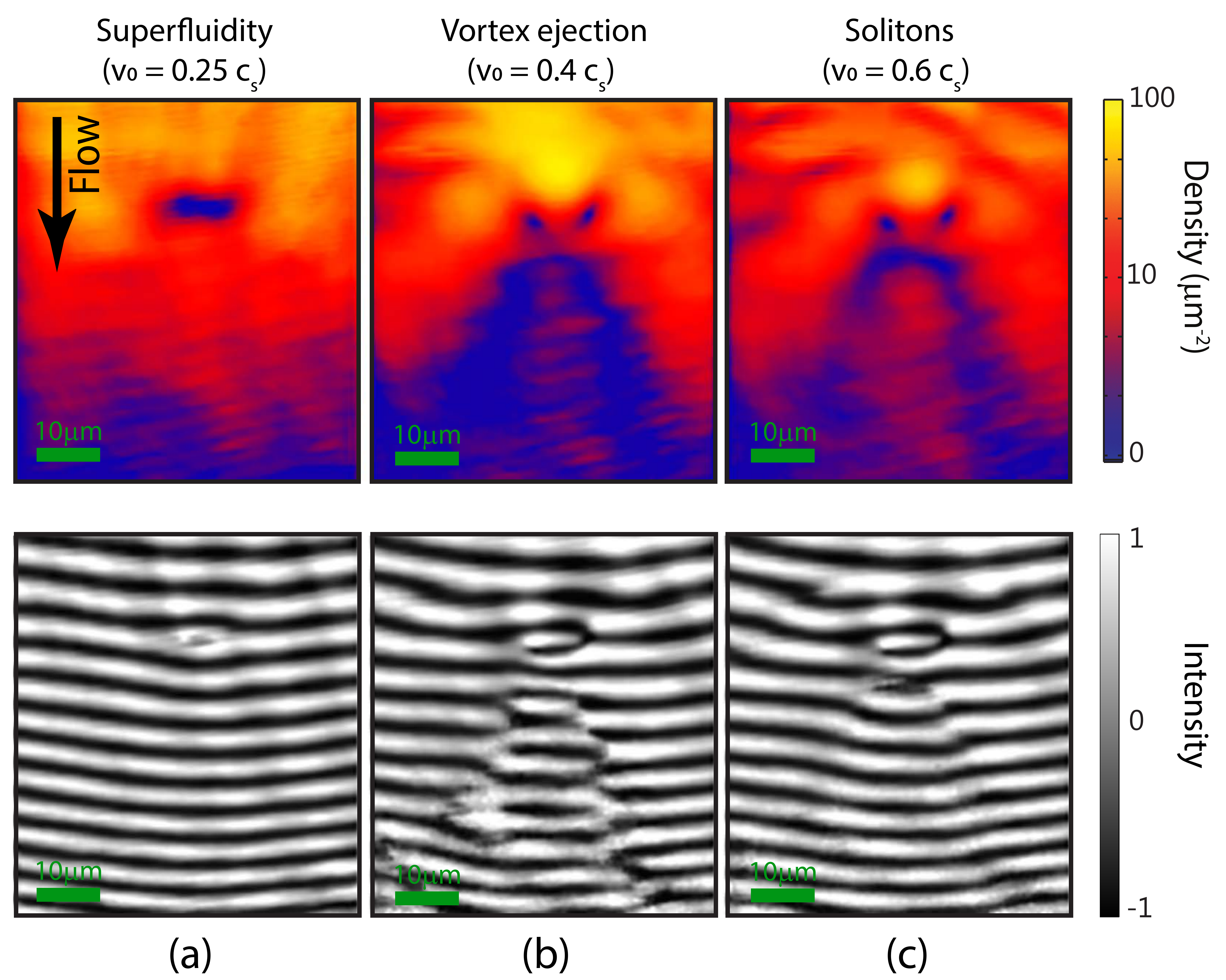}
  \caption{Top row: Images of the polariton fluid hitting a large defect. The flow is generated by a laser at $\mathbf{k} = 0.34 \mu m^{-1}$, giving a flow velocity of $0.8 \mu m/ps$. The polariton density is decreased from (a) to (c) so that the sound velocity decreases from $3.2 \mu m/ps$ (a) to $1.9 \mu m/ps$ (b) and to $1.3 \mu m/ps$ (c). Bottom row: corresponding interferograms. Superfluidity is observed in panel (a) while a turbulent pattern with phase dislocations appears in (b) and oblique dark solitons with phase jumps can be seen in (c). Reproduced with permission, from~\cite{amo2011polariton}. Copyright 2011, AAAS.}
  \label{fig5}
\end{center}
\end{figure}

Reference~\cite{amo2011polariton} reports an illustrative situation in which different hydrodynamic regimes were explored in a polariton fluid as a function of the ratio of speed of sound to fluid speed. Experiments were performed with the same sample as above with an obstacle chosen to be larger than the healing length, which is of the order of $3\mu m$ with the used polariton densities. When the fluid velocity is well below the sound velocity ($v_0 = 0.25 c_s$), the polariton ensemble is in the superfluid regime. As predicted above when the ratio of the flow velocity to the sound speed, i.e. the Mach number, is higher and reaches $v_0 / c_s = 0.4$, the fluid enters in a turbulent regime with the appearance of phase dislocations in two low-density streams in the wake of the defect. These phase dislocations can be seen in Figure \ref{fig5} in the phase map, obtained by interfering the microcavity emission with a reference beam. They correspond to the continuous emission of pairs of quantized vortices and antivortices. These vortices are topological excitations characterized by the quantized winding of the field phase from 0 to $2\pi$. Here, since the flow is inviscid, the number of vortices and antivortices is the same.

When the Mach number is further increased ($v_0 / c_s = 0.6$), the formation of oblique dark solitons is observed with a characteristic phase jump, as opposed to a stream of phase singularities. The observation of oblique solitons was surprising because they had been predicted to be affected by the snake instability~\cite{kivshar2000self}. It was actually shown in~\cite{kamchatnov2008stabilization} that the solitons remain stable because the snake instability is suppressed by the supersonic flow. Remarkably, it was shown~\cite{kamchatnov2012oblique} that the dissipation inherent to polariton fluids can further stabilize the solitons. 

Other real-time studies of this turbulence were performed in using a pulsed pump laser~\cite{sanvitto2011all,amo2009collective,nardin2011hydrodynamic, Grosso2011, Grosso2012}. After the pump is switched off, the polariton population has a characteristic decay time of order $15 ps$ during which the condensate phase is free to evolve, directly revealing the formation of vortices and antivortices.

These experiments highlight the broad variety of turbulence phenomena, akin to the ones appearing in matter, which can be simulated with polariton systems. Their all-optical control makes them highly flexible, and thus quite promising for quantum turbulence simulations. Up to recent years, a known limitation was the limited propagation length induced by the finite polariton lifetime. However, a novel method to sidestep this problem was proposed and implemented recently.

\subsection{Sustained polariton flow}

\subsubsection{A methodology relying on bistability}

Due to the dissipative character of the polaritons, their lifetime and consequently their propagation distance in the absence of pumping are limited. Typical lifetime is of order $15 ps$, which gives a maximum free propagation distance of the order of $40\mu m$ at top speed ($v \sim 2\mu m/ps$). Higher lifetimes, reaching more than $100~ps$ have been reported~\cite{Steger2013}, but they require microcavities with a very high finesse and consequently very exigent fabrication conditions. On the other hand, the presence of a resonant driving field tends to lock the phase and the density of the fluid, as mentioned above, and to  inhibit the formation of topological excitations such as vortices or solitons. Recently, an original method was theoretically proposed~\cite{pigeon2017sustained} to circumvent this problem. The idea was to use a "support" field along the propagation, with an intensity comprised within the range of the bistable regime (between the two bistability turning points).

As mentioned above, when the pump intensity is in the bistable range, the density equation for the polaritons has two solutions, shown in Figure \ref{fig1}(b). But  it has been shown that the evolution of the polariton field is quite sensitive to fluctuations. Early studies demonstrated a time-dependent hysteresis and fluctuations in this system~\cite{bonifacio1978photon,mandel1982dynamics}. More recently, the nature of the phase transitions between the two bistable states was studied theoretically~\cite{casteels2016power,casteels2017critical}, and experimentally investigated in polariton systems using artificially injected classical noise~\cite{abbaspour2015effect}. The phase transitions were also demonstrated in the presence of quantum fluctuations~\cite{rodriguez2017probing,Fink2018}. 

\begin{figure}[ht]
\begin{center}
  \includegraphics[width=1.0\linewidth]{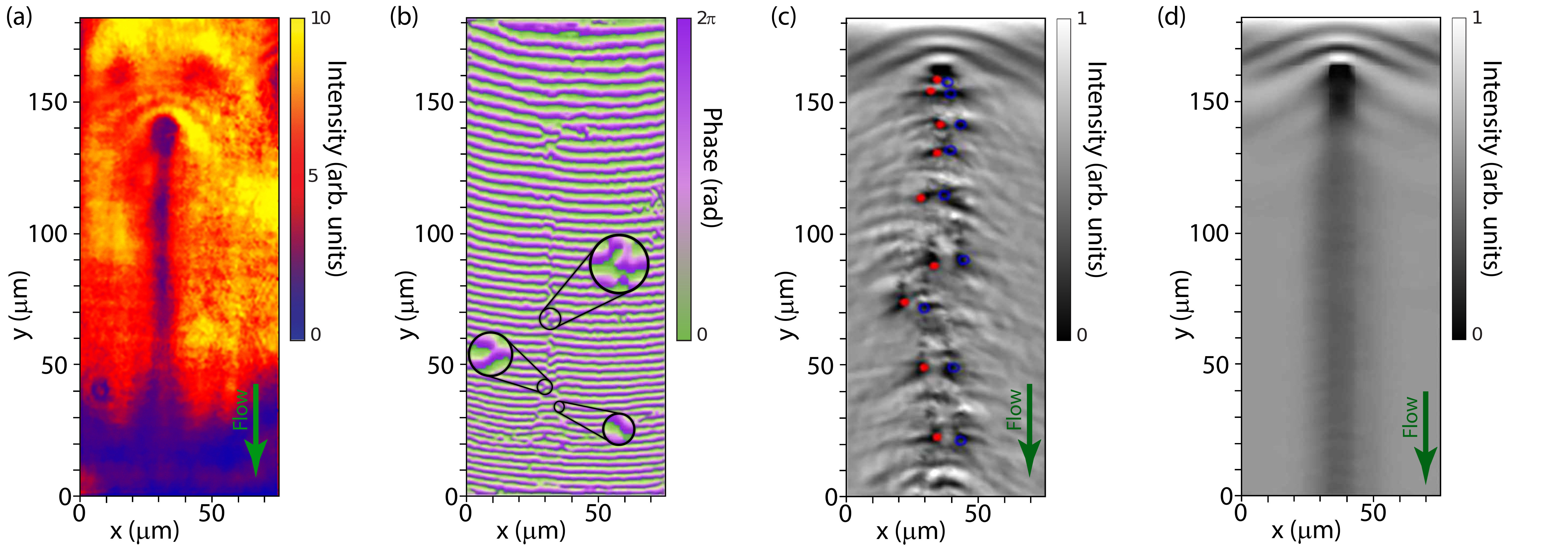}
  \caption{Vortex propagation in a polariton flow in the presence of a seed and a support beam. (a) Image of the (time-averaged) flow that hits an obstacle visible as a black dot. The seed beam is localized upstream of the obstacle, the support beam extends down to $130 \mu m$ with a large enough intensity. A dark shadow can be seen in the wake of the defect. (b) Phase diagram of the flow showing the presence of phase bifurcations (inset) characteristic of the vortices. (a-b) are from Lerario \textit{et al.}~\cite{lerario2020vortex}. (c-d) Theoretical modeling of the vortex propagation, from Pigeon \textit{et al.}~\cite{pigeon2017sustained}. (c) is a snapshot of the vortex street displaying individual vortices (blue dots) and anti-vortices (red dots), while (d) is the time-average.}
  \label{fig6}
\end{center}
\end{figure}

These results show that the two solutions in the bistable regime are sensitive to fluctuations, and that the system can easily jump from one to the other in the presence of a defect, of a potential barrier or of some interaction, avoiding the locking effect by the driving field.
Reference~\cite{pigeon2017sustained} predicted that using a support beam in the bistable regime allows to obtain a macroscopic enhancement of the propagation distance, while allowing to observe the hydrodynamics of the superfluid, including the creation of vortex streets and solitons, as shown in Figure \ref{fig6}(c). Moreover, in the case of a fluid at rest, the properties of the bistable regime allow to study the onset of the snake instabilities, as theoretically investigated in ~\cite{koniakhin2019stationary}.

\subsubsection{Sustained propagation of vortices and solitons}

The experiment was performed in a semiconductor microcavity, with two driving fields from the same laser~\cite{lerario2020vortex}. The first field is a localized seed with an intensity above the bistability threshold, which generates the superfluid polariton flow in a small region of the sample ($30\mu m$ in diameter). The second beam is the support field, with an elliptical shape and a length of about $200\mu m$. Its intensity is inside the bistability regime, so as to sustain the flow over a long distance without locking it. Both fields have the same wavevectors $|k|=0.6\mu m^{-1}$ yielding a group velocity of $0.9\mu m/ps$, with a sound velocity of $0.78\mu m/ps$ at the position of the seed beam so that the polariton flow is supersonic.

A large optical defect is then placed slightly downstream of the seed to generate Cerenkov wavefronts ahead of the defect and topological excitations in the wake of the defect, as can be seen in Figure \ref{fig6}. In the supported region these excitations can propagate on a long distance, about $120\mu m$. This distance is much longer than the maximum allowed by their lifetime ($14ps$), which would be about $40\mu m$, as can be seen in Figure \ref{fig6}.

The CCD camera image in Figure \ref{fig6}(a) shows a dark shadow in the wake of the defect. In order to analyze the phase evolution in this flow, the signal coming out of the microcavity interferes with a reference beam coming from the source laser. Figure \ref{fig6}(b) presents the disordered phase jumps visible along the dark line in the wake of the defect. As can be seen in the insets, fork-shaped patterns associated with localized vortices and antivortices are present. This is in good agreement with the theoretical predictions \cite{pigeon2017sustained}  shown in Figure \ref{fig6}(c). It can also be verified that the vortex generation rate is maximal close to the lower limit of the bistability loop, and decreases when the support intensity increases, disappearing outside the bistability regime.

More recently, the generation of a parallel dark soliton pair in the wake of a cavity structural defect was demonstrated~\cite{lerario2020parallel} when the polariton flow is further in the supersonic regime ($v =  1.52\mu m/ps,~c_s = 0.4\mu m/ps$). This bound soliton pair propagates on long distances, with a constant separation of about $8 \mu m$.

The properties of the supported vortex streams and solitons are therefore quite different from the freely propagating ones. In the latter case, the solitons are grey and the vortex streams and solitons are oblique \cite{winiecki2000vortex,kamchatnov2008stabilization}, while in the supported case they propagate along the flow direction. As mentioned above, vortex streams result from instabilities affecting solitons and possess trajectories similar to that of solitons. In the presence of a support beam, vortex streams and solitons cannot go apart because the support beam creates an  effective potential pushing them towards each other. It originates from the support beam being in phase with the field outside the streams, but out of phase with the field between the streams \cite{koniakhin2019stationary,lerario2020parallel}. On the other hand, as well known in the literature \cite{kivshar1998dark}, there is always a repulsive interaction between two dark solitons. These two opposite contributions create a minimum at finite distance in the effective potential, thus producing a bound soliton pair with a stable distance between the solitons.

One can analyze in detail the propagation dynamics of topological defects, vortices and solitons in a homogeneous superfluid on a long distance. Fine-tuning of the properties of topological excitations can be achieved by controlling the intensity of the support field, making this platform very attractive to study quantum fluid physics. This offers the opportunity to study more complex phenomena like Andreev reflections~\cite{daley2008andreev}, nucleation and trapping of vortex lattices and quantum turbulence~\cite{berloff2010turbulence}.

\subsection{Bistable dissipative solitons in 1D microcavities}

The driven-dissipative nature of polaritons provides opportunities for unveiling hydrodynamics phenomena not accessible in quantum fluids at equilibrium. The combination of bistability associated to the resonant excitation of the system and the nucleation of dark solitons in dense polariton fluids opens up the possibility of exploring bistable hydrodynamics.
\begin{figure}[ht]
\begin{center}
  \includegraphics[width=0.7\linewidth]{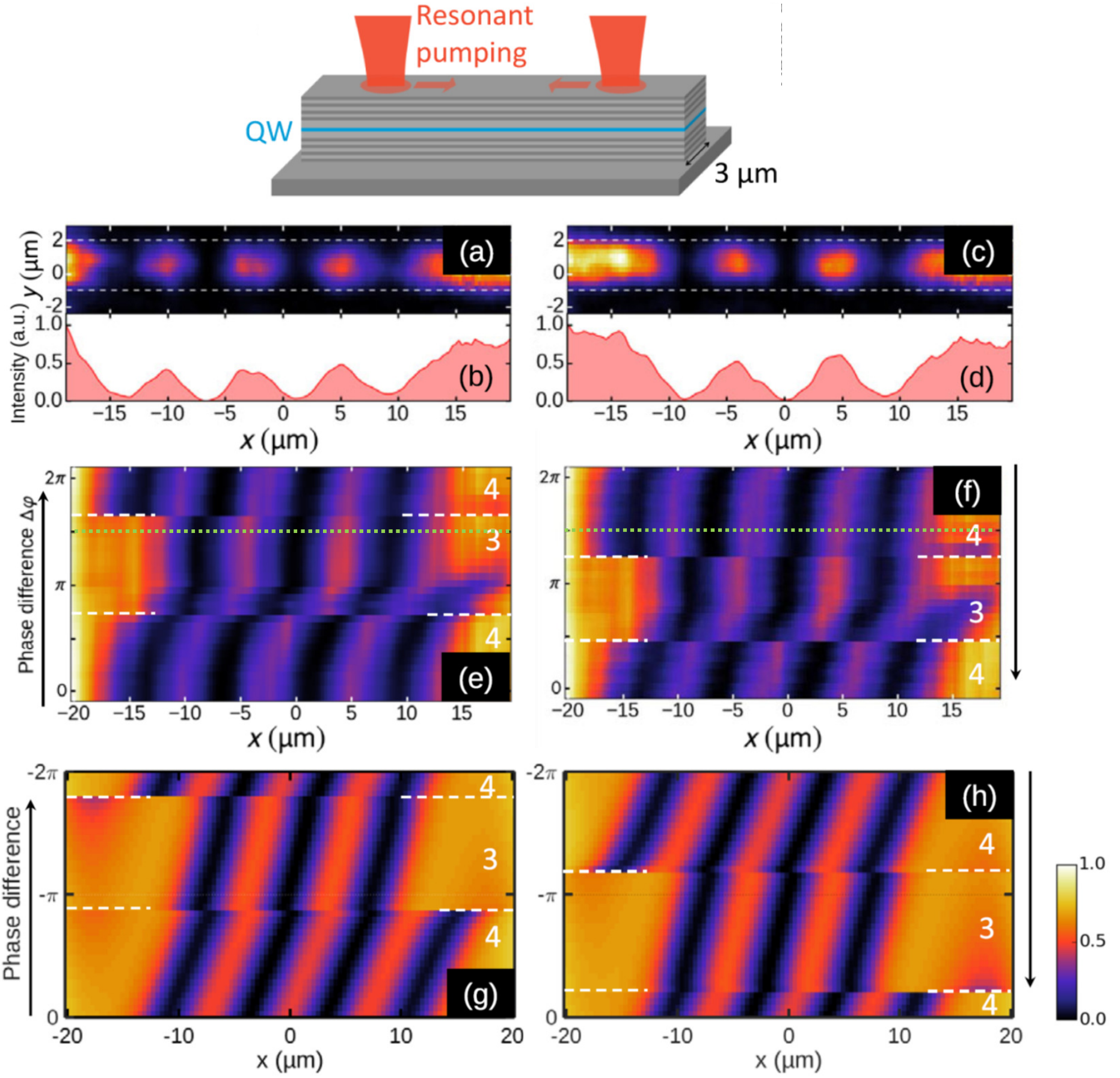}
  \caption{Phase bistability of dark solitons in 1D. (a) Real space emitted intensity in a microcavity wire in the region between the pump spots for a high injected density and a phase difference of zero between the two beams. (b) Intensity integrated vertically all along the width of the wire, evidencing the profile of four dark solitons. (c), (d), same as (a), (b), respectively, for a phase difference of $\pi$ between injection beams, displaying three dark solitons. (e), (f) Measured intensity profiles along the wire as a function of the difference of phase between the injection beams, when increasing (e) and decreasing (f) the difference. (g), (h) Calculated profiles in the conditions of (e), (f). Reproduced with permission, from~\cite{Goblot2016}. Copyright 2016, American Physical Society.}
  \label{fig:1dsolitons}
\end{center}
\end{figure}
An example of such regime was studied by Goblot and coworkers in Ref.~\cite{Goblot2016}. This work considered a one- dimensional microcavity wire fabricated via controlled etching of an AlGaAs microcavity very similar to that used in the experiments described above. The lateral confinement induced by the finite width of the wire ($3~\mu$m) results in polariton sub-bands separated in energy by several meV. The uppermost panel of Figure~\ref{fig:1dsolitons} shows the excitation scheme employed to generate dark solitons in the wire: polaritons are injected with a quasi-resonant laser in two $8~\mu$m spots separated by $50~\mu$m. The laser frequency is $0.37$~meV above the $k=0$ energy of the lowest polariton sub-band. At low excitation power, polaritons propagate away from the excitation spots and interfere in the region in between them. At high power, the polariton fluid shows the nucleation of dark solitons. Interestingly, the number of nucleated solitons depends on the phase difference between the two quasi-resonant excitation beams: for a zero phase difference, four solitons are nucleated, as shown in Figure~\ref{fig:1dsolitons}(a)-(b), while for a phase difference of $\pi$, three solitons are visible (Figure~\ref{fig:1dsolitons}(c)-(d)). Remarkably, when scanning the phase difference of the injection beams, the number of measured dark solitons shows a bi-stable behaviour. For instance, for a phase difference of $3\pi/2$, either 3 or 4 dark solitons can be nucleated between the pump spots depending on how the system got there, either by increasing the phase difference from zero or by decreasing it from $2\pi$, respectively (see green dashed line in Figure~\ref{fig:1dsolitons}(e), (f)). The experimental observations are well reproduced by numerical simulations of Equation~\ref{eq:GPE}, as shown in Figure~\ref{fig:1dsolitons}(g),~(h).

The driven-dissipative nature of polaritons allows an intuitive understanding of the observed phase bistability of dark solitons. The frequency of the injection lasers sets the local energy per polariton. This energy is shared into a kinetic and interaction energy components. The share between these two components determines the number of solitons for a given set of parameters (laser energy, separation between pumps spots, excitation density). When the phase difference between the excitation beams is modified, a phase twist is imposed (kinetic energy), which alters the balance of kinetic and interaction energies and results in a change in the number of solitons. A nonlinear feedback mechanism similar to that described in Sec.~\ref{sec:bistability} results in the observed bi-stable behaviour.

\section{Multistable dissipative gap solitons in a 1D flat band}
Another type of nonlinear excitations that can be resonantly driven in a polariton fluid are bright solitons.
They correspond to a high intensity region whose spatial stability is governed by the balance between a focusing and a defocusing mechanism. In atomic BEC, the dispersive spreading induced by the kinetic energy of atoms can be balanced by attractive interatomic interactions \cite{Khaykovich1290, Strecker2002}. 
In the case of repulsive interaction like in a polariton fluid, bright solitons are stable when driven in the vicinity of negative mass bands \cite{Egorov2009}. They have been demonstrated when driving cavity polaritons beyond the point of inflection of the lower polariton band, where the polariton mass is negative \cite{Sich2011soliton}.

Another configuration to generate bright solitons is obtained when driving the system in a gap.
To design a band structure for polaritons with controlled mini-bands and gaps, one can sculpt the microcavity on micrometric scales (which is possible with current nanotechnology) and fabricate lattices of coupled micropillars \cite{Schneider_2016}.

\begin{figure}[ht]
\begin{center}
  \includegraphics[width=1 \linewidth]{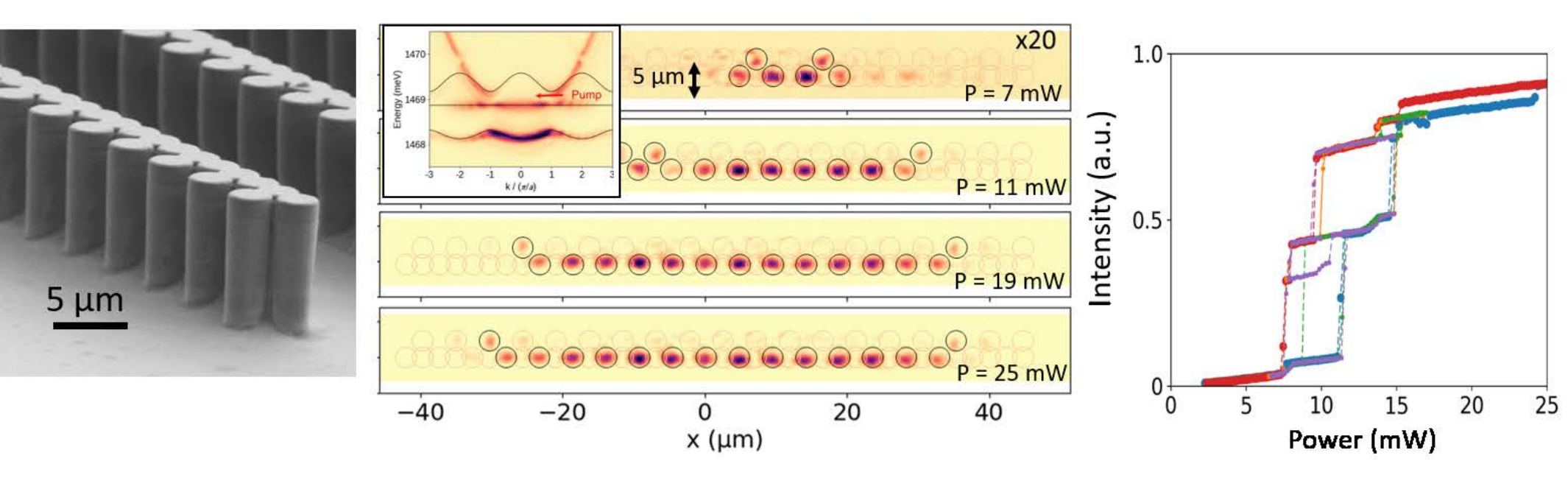}
  \caption{Multistability of gap solitons in a 1D flatband: (Left): Scanning electron microscopy image of a 1D Lieb lattice of coupled micropillar cavities; (Center): inset: Photoluminescence emission as a function of energy and wave-vector revealing a band structure with a central flat band; Spatially resolved transmission of the quasi-resonant pump laser for four different excitation power; (right): Total transmitted intensity when scanning the power up and down. Adapted with permission from ~\cite{Goblot2019}. Copyright 2016, American Physical Society.}
  \label{fig:gapsolitons}
\end{center}
\end{figure}

Individual modes of each resonator hybridize and form bands with properties related to the geometry and topology of the lattice \cite{AMO-CRAS}.
A particularly interesting lattice to explore bright gap solitons is the 1D Lieb lattice \cite{Lieb1989} as shown in Figure \ref{fig:gapsolitons} (left). It contains three sites per unit cell and a geometric frustration which results in a gaped flat band, with a very large degeneracy of localized eigenstates \cite{Leykam2018}.
Such a flat band emulates the physics of infinitely massive particles. Because of the absence of any kinetic energy, the effect of interactions is particularly strong in a flat band and emerges as soon as interactions overcome the polariton linewidth. 
This physics is illustrated in Figure \ref{fig:gapsolitons} (center): a 1D polariton Lieb lattice is quasi-resonantly driven by a laser, which is  blue detuned by $90 \mu eV$ with respect to the flatband and shaped in a gaussian spot elongated along the lattice direction (diameter of about $40 \mu$m) \cite{Goblot2019}.
When the laser power is increased, the  intensity transmitted through the lattice increases by consecutive abrupt jumps, separated by plateaus where the intensity barely varies.
Resolving the transmitted signal in real space, one observes that each plateau corresponds to the formation of a non-linear domain extending over an integer number of unit cells, and presenting very abrupt edges (see Figure \ref{fig:gapsolitons} center).
They are formed by an integer number of localized eigenstates (also named plaquettes), for which the local interaction energy $\hbar g$ overcomes the laser detuning. 
This condition is realized first at the center of the excitation spot (where the excitation power is the strongest) and is then fulfilled for an increasing number of plaquettes as the excitation is increased.
This explains why the size of the gap solitons increases by abrupt jumps every time additional plaquettes switch into the non-linear regime. Outside the gap solitons, the interaction energy can neither be accommodated by the system, nor converted into kinetic energy.
This is why the gap solitons are discrete and present very abrupt edges. Interestingly, because of the driven dissipative nature of the polariton fluid, each plaquette can undergo bistability. Scanning the power up and down around each transmission plateau reveals striking multistability  behaviour (see Figure \ref{fig:gapsolitons} right) with many intricate hysteresis cycles.
This is a direct consequence of both the high degeneracy of the flat band and the driven dissipative nature of the system. This multistability is a very distinct behavior as compared to conservative systems.
Lattices presenting flatbands are particularly interesting for the generation of strongly correlated many body phases, which could emerge in polariton lattices with stronger interactions \cite{Wu2007,Sun2011,Neupert2011}. 

\section{Analogue gravity with polaritons}
The hydrodynamics of polaritons in microcavities can be engineered to study effects of analogue gravity \cite{volovikUniverseHeliumDroplet2003,barceloAnalogueGravity2011a,jacquetNextGenerationAnalogue2020}.
This consists in the laboratory study of phenomena connected  with  scalar quantum fields on curved spacetimes.
Via such experiments, we gain access to  effects  of  the  interplay  between  general  relativity  and quantum physics that normally elude detection: typically, the amplification of waves in regimes of extreme curvature.
The most famous of these phenomena undoubtedly is the Hawking effect --- the amplification of quantum vacuum fluctuations at the event horizon of static black holes.
This results in the emission of particles pairs : the Hawking radiation that escapes from the black hole and its partner that falls inside the horizon \cite{hawking_black_1974,hawking_particle_1975}.

The possibility of observing the Hawking effect has driven the development of analogue gravity experiments since the turn of the 21st century \cite{unruh_experimental_1981} and established systems now range from water tanks \cite{rousseaux_observation_2008,weinfurtner_measurement_2011,euveObservationNoiseCorrelated2016,euveScatteringCoCurrentSurface2020a} to optical fibers and crystals \cite{philbin_fiber-optical_2008,faccio_analogue_2010,drori_observation_2019} or ion traps \cite{wittemerPhononPairCreation2019a}.
In this context, BECs \cite{lahav_realization_2010,jaskulaAcousticAnalogDynamical2012,eckelRapidlyExpandingBoseEinstein2018a} are an important platform as experimental techniques for quantum measurements are well developed and accompanied by a rich theoretical description of entangled phonon pairs emission in various scenarios.
In fact, the observation of spontaneous emission at the horizon has so far only been reported in a series of experiments in an atomic BEC \cite{steinhauerObservationQuantumHawking2016a,munoz_de_nova_observation_2019,kolobovSpontaneousHawkingRadiation2019}.

Sonic horizons can also be realized with polaritons in microcavities. Let us consider a one-dimensional polariton flow. The wave equation of Bogoliubov excitations (sound waves at low $k$) in the flow is obtained by linearizing Equation \eqref{eq:hydroGPE} and Equation (18) around a background state by the Bogoliubov method (see Equation \eqref{eq:perturbedwf}) and neglecting the quantum pressure:
\begin{equation}
    \label{eq:waveeqsound}
\frac{\rho_0}{c_s^2}\left(\left(c_s^2-v^2\right)\partial x^2-\partial t^2-2v\partial t\partial x\right)\delta\psi=0
\end{equation}
with $\delta\psi=\frac{\delta\rho}{\rho}$, $c_s$ the local speed of sound and $v$ the flow velocity of the polariton fluid (as defined under Equation \eqref{eq:hydroGPE}).
Equation \eqref{eq:waveeqsound} is isomorphic to the d'Alembertian $  \Delta\psi_1\equiv\frac{1}{\sqrt{-\eta}}\partial_\mu\left(\sqrt{-\eta}\eta^{\mu\nu}\partial_\nu\delta\psi\right)=0$ of scalar waves on a 1+1D spacetime of metric $\eta_{\mu\nu}$~\cite{MaximeFootnote}

 \begin{equation}
     \label{eq:lineelement}
         ds^2\equiv\eta_{\mu\nu}\partial^\mu \partial^\nu=\frac{\rho_0}{c_s}\left(-\left(c_s^2-v^2\right)dt^2+dx^2-2vdtdx\right).
 \end{equation}
There is a  horizon in the spacetime of Equation \eqref{eq:lineelement} when $c_s=v$ \cite{unruh_experimental_1981}.
Interestingly, such a geometry naturally occurs in polariton flows pumped with a beam of arbitrary shape, \textit{e.g.} a Gaussian beam, simply because the density of polaritons will decrease away from the center of the pump.
As the density of polaritons decreases, so will $c_s$.
In contrast, $v$ will increase due to the interaction energy being converted to kinetic energy.
If at the pump center $c_s>v$ (subsonic flow), there will come a point in the cavity where $c_s=v$ and, beyond that point, $v>c_s$ (the flow is supersonic). Spontaneous emission of phonon pairs from the vacuum is expected to occur at sonic horizons in polaritons \cite{solnyshkovBlackHolesWormholes2011}.

If the horizon is steep enough, the temperature of spontaneous emission should permit detection \cite{gerace_analog_2012}.
The first experimental demonstration of a polaritonic flow featuring a horizon was made in 2015 \cite{Nguyen2015}.
There the horizon was made steep by etching a defect in the cavity, which modifies the density of polaritons in a non-adiabatic way.
The horizon may also be steepened by means of an optical defect \cite{jacquetPolaritonFluidsAnalogue2020}.
As in the atomic BEC, paired emission is characterised by a correlation pattern in density fluctuations across the horizon \cite{Grisins2016}.
Well-established quantum optical techniques such as homodyne detection for the measurement of \textit{e.g.} squeezing \cite{Boulier_squeezing_2014} could be used to quantify the entanglement content \cite{busch_spectrum_2014} of these correlations and establish their vacuum fluctuation origins.

It is also possible to create rotating polariton flows akin to the Kerr black hole \cite{Marino_proposal_2008} by pumping the polaritons with a Laguerre-Gauss beam of arbitrary degree of optical angular momentum, as in the experiments of \cite{jacquetPolaritonFluidsAnalogue2020}.
Rotating flows thus engineered feature two important surfaces: an inner horizon and an ergosurface \cite{visser_acoustic_1998}, where the radial and total velocity of the flow become supersonic, respectively.
Such a setup could be used to observe the amplification of incoming waves by rotational superradiant scattering, as was done with water waves in \cite{torres_rotational_2017}.
Furthermore, rotating geometries are the perfect test bench for effects beyond the amplification of small amplitude waves.
For example, one could impinge a pair of vortex-anti vortex on the ergosurface and extract energy from it by the Penrose effect \cite{Solnyshkov2019}.
The propagation of solitons on a vortex-type flow could also be used to probe the kinematics in this geometry.

Here again, the driven-dissipative nature of microcavity polaritons is an advantage: it provides an energy drain (photons escaping the cavity), which is an essential ingredient in the study of phenomena associated with the physics of rotating black holes.
Thus, polaritons in microcavities, like other fluids of light \cite{vockeRotatingBlackHole2018b,prainSuperradiantScatteringFluids2019}, are a versatile platform for the experimental study of effects of analogue gravity.

\section{Conclusion and further perspectives}
We have seen that polariton microcavities provide a versatile platform for the study of quantum hydrodynamics. The significant strength of the polariton-polariton interactions, the direct optical control of the injected fluid and the direct access to the phase and amplitude information of the polariton field using standard optics setups have been major assets for the study of superfluidity and the nucleation of vortices and solitons~\cite{Carusotto2013}.
Far from being a limitation, the finite polariton lifetime opens the possibility of exploring phenomena that goes beyond what is expected in equilibrium systems.
Three examples of such situation have been described above (sustain hydrodynamics in a two-dimensional flow, phase bi-stability in one-dimensional dark solitons, and multistability in flat bands). Open questions include the formation of turbulent cascades in the presence of dissipation at all length scales~\cite{Koniakhin2020}, as it is the case for polaritons, and the validity of the entropy arguments explaining the formation of inverse cascades in lossless two dimensional systems~\cite{Simula2014, Gauthier2019, Johnstone2019}.


Beyond this kind of problems, which are inspired by phenomena predicted and/or observed in equilibrium systems, genuinely out of equilibrium physics is at hand in this platform. An example is the study of the Kardar-Parisi-Zhang universality class, which might be accessible in polariton systems via the decay of the first order coherence under non-resonant excitation in one dimensional microcavities~\cite{Gladilin2014, Ji2015, He2015,Squizzato2018}. The observation of these universality classes in polaritons would open the door to the study of such universal behaviour in two dimensions, which is yet to be revealed.

\medskip
\textbf{Acknowledgments} \par
We acknowledge the support of the ANR project "Quantum Fluids of Light" (ANR-16-CE30-0021). This work has received funding from the European Union’s Horizon 2020 research and innovation program under grant agreement No. 820392 (PhoQuS). A.B. and Q.G. are members of the Institut Universitaire de France (IUF).

\medskip
\bibliographystyle{unsrt}
\bibliography{Biblio.bib}

\begin{thebibliography}{100}
\bibitem{MaximeFootnote}
In the covariant notation, the upper and lower greek indices are indices of
  coordinates, used for the time and space components (in this order, ranging
  over the indexing set $\{0,1\}$, equivalent to the traditional $\{t,x\}$).
  $\eta_{\mu\nu}$ and $\eta^{\mu\nu}$ are the covariant and contravariant
  metric, respectively.
\end{thebibliography}

\end{document}